\documentstyle[aps]{revtex}

\begin{document}
\draft
\preprint{KANAZAWA 93-10}
\title{
Monopole Condensation and Confinement in SU(2) QCD (2)
}
\author{
Hiroshi Shiba\cite{sh} and Tsuneo Suzuki\cite{su}}
\address{
Department of Physics, Kanazawa University, Kanazawa 920-11, Japan
}
\date{\today}
\maketitle
\begin{abstract}
Monopole and photon contributions to Wilson loops are calculated using 
Monte-Carlo simulations of SU(2) QCD in the maximally abelian gauge. 
The string tensions of SU(2) QCD are well reproduced by extended 
monopole contributions alone. 
\end{abstract}

\narrowtext
To understand color confinement mechanism in QCD is absolutely 
necessary
for us to analytically explain hadron physics starting from QCD
\cite{thooft1,mandel}.
The 'tHooft idea of abelian projection in which 
a partial gauge-fixing is done keeping the maximal abelian 
torus group unbroken 
is very interesting\cite{thooft2}.
Then QCD can be regarded as a $U(1) \times 
U(1)$ abelian gauge theory with magnetic monopoles and electric charges.
An interesting gauge has been found among infinite ways of 
gauge-fixing  for the abelian projection.
It is called  maximally abelian (MA) 
gauge\cite{kron,yotsu,hioki,suzu1} 
in which link gauge fields
are forced to become abelian as much as possible. 

In the preceding note, the authors\cite{shiba} have shown 
in the MA gauge 
and in $SU(2)$ QCD 
that entropy dominance over energy of the monopole loops, i.e.,
condensation of the monopole loops occurs 
in the confinement phase if extended 
monopoles\cite{ivanenko} are considered. 
After the abelian projection in the MA gauge, infrared behaviors of 
$SU(2)$ QCD seem to be described by a compact-QED like $U(1)$ theory 
with the running
coupling constant instead of the bare one and with 
the monopole mass on a 
dual lattice.
The confinement in $SU(2)$ QCD may be interpreted 
as the (dual) Meissner 
effect due to the abelian monopole condensation.

If the monopoles alone are responsible for the confinement mechanism, 
the string tension which is a key quantity 
of confinement must be explained 
by monopole contributions. This is realized in compact QED as shown 
recently by Stack and Wensley\cite{stack}. 
The aim of this note is to show 
that the same thing happens also in $SU(2)$ QCD 
by means of evaluating 
monopole and photon contributions to Wilson loops.

After the abelian projection in the MA gauge, 
a diagonal matrix $u(s,\mu)$ can be extracted 
uniquely from the original
$SU(2)$ link field. 
The diagonal matrix $u(s,\mu)$ corresponds
to a $U(1)$ gauge field written by an angle 
variable $\theta_{\mu}(s)$.

Now we show an abelian Wilson loop operator 
(which we consider after the 
abelian projection) is 
rewritten  by a product of monopole and photon contributions. Here
we take into account only a simple Wilson loop, say, of size
$I \times J$. Then such an 
abelian Wilson loop operator is expressed as 
\begin{eqnarray}
W = \exp\{i\sum J_{\mu}(s)\theta_{\mu}(s)\}, 
\end{eqnarray}
where  $J_{\mu}(s)$ 
is an external current taking $\pm 1$ along the Wilson loop. 
Since $J_{\mu}(s)$
is conserved, it is rewritten for such a simple Wilson loop 
in terms of an antisymmetric  
variable $M_{\mu\nu}(s)$ 
as $J_{\nu}(s)=\partial_{\mu}'M_{\mu\nu}(s)$, 
where $\partial'$ is a backward 
derivative on a lattice. 
$M_{\mu\nu}(s)$ takes $\pm 1$ on a surface with the 
Wilson loop boundary.
Although we can choose any surface of such a type, 
we adopt a minimal 
surface here.  
We get 
\begin{equation}
W  =  \exp \{-\frac{i}{2}\sum M_{\mu\nu}(s)f_{\mu\nu}(s)\}
\label{W},
\end{equation}
where $f_{\mu\nu}(s)= \partial_{\mu}\theta_{\nu}(s) - 
\partial_{\nu}\theta_{\mu}(s)$ and $\partial_{\mu}$ is a 
forward derivative on a lattice.
The gauge plaquette variable can be decomposed into $f_{\mu\nu}(s) = 
\bar{f}_{\mu\nu}(s)+ 2\pi n_{\mu\nu}(s)\ $  where 
$\bar{f}_{\mu\nu}(s) \in [-\pi, \pi] $ corresponds to a 
field strength and $n_{\mu\nu}(s)$ 
is an integer-valued plaquette variable\cite{comment1}
denoting the Dirac string. Since 
$M_{\mu\nu}(s)$ and $n_{\mu\nu}(s)$ are integers, the latter does not 
contribute to Eq.\ (\ref{W}). Hence $f_{\mu\nu}(s)$ in 
Eq.\ (\ref{W}) is replaced by
$\bar{f}_{\mu\nu}(s)$. 
Using a decomposition rule 
\begin{eqnarray*}
M_{\mu\nu}(s) = -\sum D(s-s')[\partial'_{\alpha}(\partial_{\mu}
M_{\alpha\nu}-\partial_{\nu}M_{\alpha\mu})(s')\\
+ \frac{1}{2}\epsilon_{\alpha\beta\mu\nu}
\epsilon_{\alpha '\beta\rho\sigma}
\partial'_{\alpha}\partial_{\alpha '}M_{\rho\sigma}(s')],
\end{eqnarray*}
we get 
\begin{eqnarray}
W\    & = & W_{1} \cdot W_{2} \label{w12}\\
W_{1} & = & \exp\{-i\sum \partial'_{\mu}\bar{f}_{\mu\nu}(s)
D(s-s')J_{\nu}(s')\} \nonumber \\
W_{2} & = & \exp\{2\pi i\sum k_{\beta}(s)D(s-s')\frac{1}{2}
\epsilon_{\alpha\beta\rho\sigma}\partial_{\alpha}M_{\rho\sigma}(s')\}, 
\nonumber
\end{eqnarray}
where a monopole current $k_{\mu}(s)$ is defined as $k_{\mu}(s)= 
(1/4\pi)\epsilon_{\mu\alpha\beta\gamma}\partial_{\alpha}
\bar{f}_{\beta\gamma}(s)$ following DeGrand-Toussaint\cite{degrand}.
$D(s)$ is the lattice Coulomb propagator. 
Since $\bar{f}_{\mu\nu}(s)$ corresponds to the field strength of the 
photon field, 
$W_{1} (W_{2})$ 
is the photon (the monopole) contribution to the Wilson loop.

To study the features of both contributions, 
we evaluate the expectation 
values $\langle W_1 \rangle$
and $\langle W_2 \rangle$ separately and compare them with those 
of $W$. 

The Monte-Carlo simulations were done on $24^4$ lattice from $\beta =
2.4$ to $\beta =2.8$. All measurements were done every 30 sweeps after
a thermalization of 1500 sweeps. We took 50 
configurations totally for
measurements. The gauge-fixing criterion 
is the same  as done in Ref.\ \cite
{ohno}. Using gauge-fixed configurations, we evaluated monopole 
currents. As shown in the previous note\cite{shiba}, type-2 extended
monopole 
loops with $b > b_c \sim 5.2\times 10^{-3}(\Lambda_L)^{-1}$ 
condense, where $b=na(\beta)$ for $n^3$ extended monopoles 
and $a(\beta)$ is the lattice constant. 
So we measured $2^3$ extended 
monopole with $b=2a(\beta)$ of the type-2\cite{ivanenko}. 
Then the effective (renormalized)
lattice volume becomes $12^4$.
Since the original lattice is $24^4$, 
$2^3$ extended monopoles are the 
largest from which we can get useful data of the static 
potentials from Wilson loops. For $\beta =2.7$ and $2.8$,
the value $b=2a(\beta)$ becomes less than $b_c$ and so the 
monopoles  may not reproduce the string tension. 

We have evaluated the averages of $W$ using abelian 
link variables  (called abelian), of 
$W_1 \cdot W_2$ (called total), of $W_1$ (photon part),  
and $W_2$ (monopole 
part), separately. Both the first and the second averages are 
evaluated to check reliability of the data, since both 
should be equivalent 
as known from Eq.\ (\ref{w12}). 

The results are summarized as follows.
\begin{enumerate}
\item
The monopole contributions to Wilson loops are 
obtained with  relatively 
small errors. Surprisingly enough, 
the Creutz ratios of the monopole contributions 
are almost independent of the loop size as shown 
partially in Table \ref{table1}.
This means that the monopole contributions are composed  only 
of an area, a perimeter, and a constant terms. 
\item
Assuming the static potential is given 
by a linear + Coulomb + constant
terms, we can determine them from the 
least square fit to the Wilson loops
\cite{itoh}. We plot their data in Fig.\ \ref{pot25} 
(at $\beta =2.5$) and in 
Fig.\ \ref{pot26} (at $\beta =2.6$). 
We find the monopole contributions are responsible for
the linear-rising behaviors. The photon part contributes only to the 
short-ranged region. 
There seems to exist a small discrepancy between the 
abelian and the monopole + photon parts for $R/a =12$, 
but  finite-size effects are expected there.  
Similar data are obtained for $\beta =2.4$.
\item
This is seen more clearly from the data of the string tensions 
which are determined from the static potentials. They are shown in
Fig.\ \ref{sigma}. Systematic errors 
coming from various least square fits are 
not completely certain and are 
not plotted in the figure, although they are not negligible.
The string tensions are well reproduced by the 
monopoles alone for $\beta \le
2.6$ and the photon part has almost vanishing string tensions.
The string tensions  evaluated from 
the total part (which are not shown here)  
are  consistent with those of abelian 
and monopoles. 
\item
At $\beta =2.7$ and $2.8$, the monopoles which have  $b<b_c $ 
do not seem to reproduce the abelian 
string tensions as shown in Table \ref{table2}. 
However the string tensions from the total part which 
should be equal to the abelian ones are also smaller.
The origin of the 
difference resides in the smallness of the 
renormalized lattice volume.
To check it, we have adopted only even-sized 
abelian Wilson loops which 
correspond to the total ones with the 
lattice spacing $2a(\beta)$ and made 
the least-square fit. The string tensions are almost 
unchanged for $\beta=2.4, 2.5$, and $2.6$ (see Fig.\ \ref{sigma}), 
but they become smaller for $\beta= 2.7$ and $2.8$ 
with large errors. 
The difference between the abelian (even only) 
and the total becomes smaller. 
In conclusion, to get definite results for $\beta \geq 2.7$ 
in this framework of the analysis, we have to adopt much 
larger lattices like $48^4$.
We can not conclude at present that the monopoles with 
$b>b_c$ alone can reproduce the string tensions as is expected. 
\item
We have derived also  Coulomb coefficients from the static potentials 
as shown in Fig.\ \ref{coulomb}. 
The monopole part has almost vanishing Coulomb 
coefficients which is in agreement with the constant behaviors of 
the Creutz rations of the monopole part as 
shown above. The photon part has 
large coefficients, but they do not reproduce the coefficients of 
the abelian static potentials.
But again they are not far from the Coulomb 
coefficients from  Eq.\ (\ref{w12})
(called total in Fig.\ \ref{coulomb}) and those from 
even-sized abelian Wilson loops.
The following may be interesting.
The photon parts are evaluated on an effective 
lattice with $b=2a(\beta)$.
Hence they have different values of $b$ for differnt $\beta$.
The Coulomb coefficients of the photon parts are well reproduced 
by the $SU(2)$ running coupling constants $g(b)$ with $b=2a(\beta)$, 
i.e., $-g(b)^2/16\pi$, where 
\begin{eqnarray}
g^{-2}(b)= \frac{11}{24\pi^2}\ln(\frac{1}{b^2\Lambda^2}) 
+ \frac{17}{44\pi^2}
\ln\ln(\frac{1}{b^2\Lambda^2}).
\end{eqnarray}
The scale parameter $\Lambda$ determined is 
$\Lambda \sim 46\Lambda_L $ 
which is 
quite near the value $\Lambda \sim 42\Lambda_L$ 
fixed from the monopole action.
\end{enumerate}

Our analyses in this note as well as in the previous 
one\cite{shiba} show 
that abelian monopoles are responsible for 
confinement in $SU(2)$ QCD 
and condensation of the monopoles is the confinement mechanism 
if the abelian projection is done in the MA gauge.

Finally we make some comments on the results of both notes.
\begin{enumerate}
\item
Why is the MA gauge so nice? 
Note that an abelian projection reduces QCD 
into an abelian theory with 
diagonal gluons as a photon-like particle and 
off-diagonal gluons  as charged partices.  
The characteristic features of the 
MA gauge among many abelian projections are in the following.
The MA gauge is defined in such a way that a quantity
\begin{eqnarray*}
R= \sum_{s,\mu} \{U_1(s,\mu)^2 + U_2(s,\mu)^2\}
\end{eqnarray*}
is minimized, where $U_1(s,\mu)$ and $U_2(s,\mu)$ 
are components of a $SU(2)$ 
link field $U(s,\mu)=U_0(s,\mu)+\vec{U}(s,\mu)\cdot\vec{\sigma}$.
Hence the gauge condition forces as many link fields as 
possible to become abelian, i.e., diagonal. 
However one can not let all link 
fields diagonal, i.e., $R=0$ as seen from a histogram analysis shown 
in Ref.\ \cite{suzu2} and from the correlation of $R$ with the monopole 
density\cite{ohno}.
If we see only long-ranged physics, 
$SU(2)$ QCD in the MA gauge may be regarded as 
a $U(1)$ theory with abelian link photon fields and abelian monopoles. 

Is there any other gauge showing similar 
behaviors?  There seem to exist many candidates. 
For example, the following 
gauge may be interesting in which 
\begin{eqnarray*}
R= \sum_{s,\mu} \{U_1(s,\mu)^2 + U_2(s,\mu)^2\}^n \hspace{5mm} (n>0)
\end{eqnarray*}
is minimized. The condition also forces link fields to be diagonal as 
much as possible.  The case $n=2$ leads us to 
$\sum_{s,\mu} A^+_{\mu}(s)A^-_{\mu}(s)
(\partial_{\mu} \pm igA^3_{\mu}(s))A^{\pm}_{\mu}(s) =0$ 
in the continuum limit. The study in the gauge is in progress.

\item
Using the method developed in Refs.\ \cite{samuel,stone}, the abelian 
monopole action derived in \cite{shiba} may be mapped on to a field 
theoretical model of an abelian 
(dual) Higgs system. The model is just 
equal to a Ginzburg-Landau type theory 
which one of the authors (T.S.) derived earlier 
assuming abelian dominance and
monopole condensation\cite{suzuki}. 
\item
To extend our method to a $T \neq 0$ system and also to
$SU(3)$ with or without dynamical quarks is  
very interesting. 
These studies are also in progress.
\end{enumerate}

We wish to acknowledge Yoshimi Matsubara 
for useful discussions especially on 
the entropy decreasing effects of the monopole loops.
This work is financially supported by JSPS 
Grant-in Aid for Scientific  
Research (c)(No.04640289).

\begin{table}
\caption{Creutz ratios from abelian and monopole 
Wilson loops at $\beta =2.6$. 
The Monopole Creutz ratio values are devided by 
$4$, being adjusted to those 
in unit $a(\beta)$. \label{table1}
}
\begin{tabular}{lcc}
Creutz ratios & abelian & monopole \\
\tableline
$\chi (2,2)  $ &  0.0872(2)    &          \\
$\chi (3,3)  $ &  0.0460(3)    &          \\
$\chi (4,4)  $ &  0.0323(4)    & 0.0173(2)\\
$\chi (5,5)  $ &  0.0270(7)    &          \\
$\chi (6,6)  $ &  0.0248(9)    & 0.0169(3)\\
$\chi (7,7)  $ &  0.0255(13)   &          \\
$\chi (8,8)  $ &  0.0196(25)   & 0.0172(4)\\
$\chi (9,9)  $ &  0.0147(66)   &          \\
$\chi (10,10)$ &  0.0307(160)  & 0.0178(6)\\
$\chi (11,11)$ &  undeterminable &    \\
$\chi (12,12)$ &  undeterminable & 0.0178(9)\\
\end{tabular}
\end{table}

\begin{table}
\caption{String tensions $\sigma/\Lambda^2_L$ at 
$\beta =2.7$ and $2.8$. 
Abelian and total mean 
those from Wilson loops evaluated using usual 
link fields and Eq. (3) 
respectively. Abelian (even) means the least-square fit using 
only even-sized abelian Wilson loops.
\label{table2}}
\begin{tabular}{lccccc}
$\beta$ & abelian & abelian (even) & total & monopole & photon\\
\tableline
2.7 & 1962(37) & 1467(220) & 1238(516)& 1087(394)& -36(62) \\
2.8 & 2130(32) & 1621(187) & 1269(326)& 1025(261)& -21(134)\\
\end{tabular}
\end{table}

\begin{figure}
\caption{
Static potentials $aV(R)$ versus $R/a$ at $\beta =2.5$.
The values are shifted by a constant.}
\label{pot25}
\end{figure}

\begin{figure}
\caption{
Static potentials $aV(R)$ versus $R/a$ at $\beta =2.6$.
The values are shifted by a constant.}
\label{pot26}
\end{figure}

\begin{figure}
\caption{
String tensions at $\beta =2.4, 2.5$, and $2.6$.}
\label{sigma}
\end{figure}

\begin{figure}
\caption{
Coulomb coefficients.}
\label{coulomb}
\end{figure}


\begin{references}
\bibitem{comment1} This condition 
$\bar{f}_{\mu\nu}(s) \in [-\pi, \pi] $
is applicable only in the case of
defining a smallest $1^3$ monopole.
\end{references}
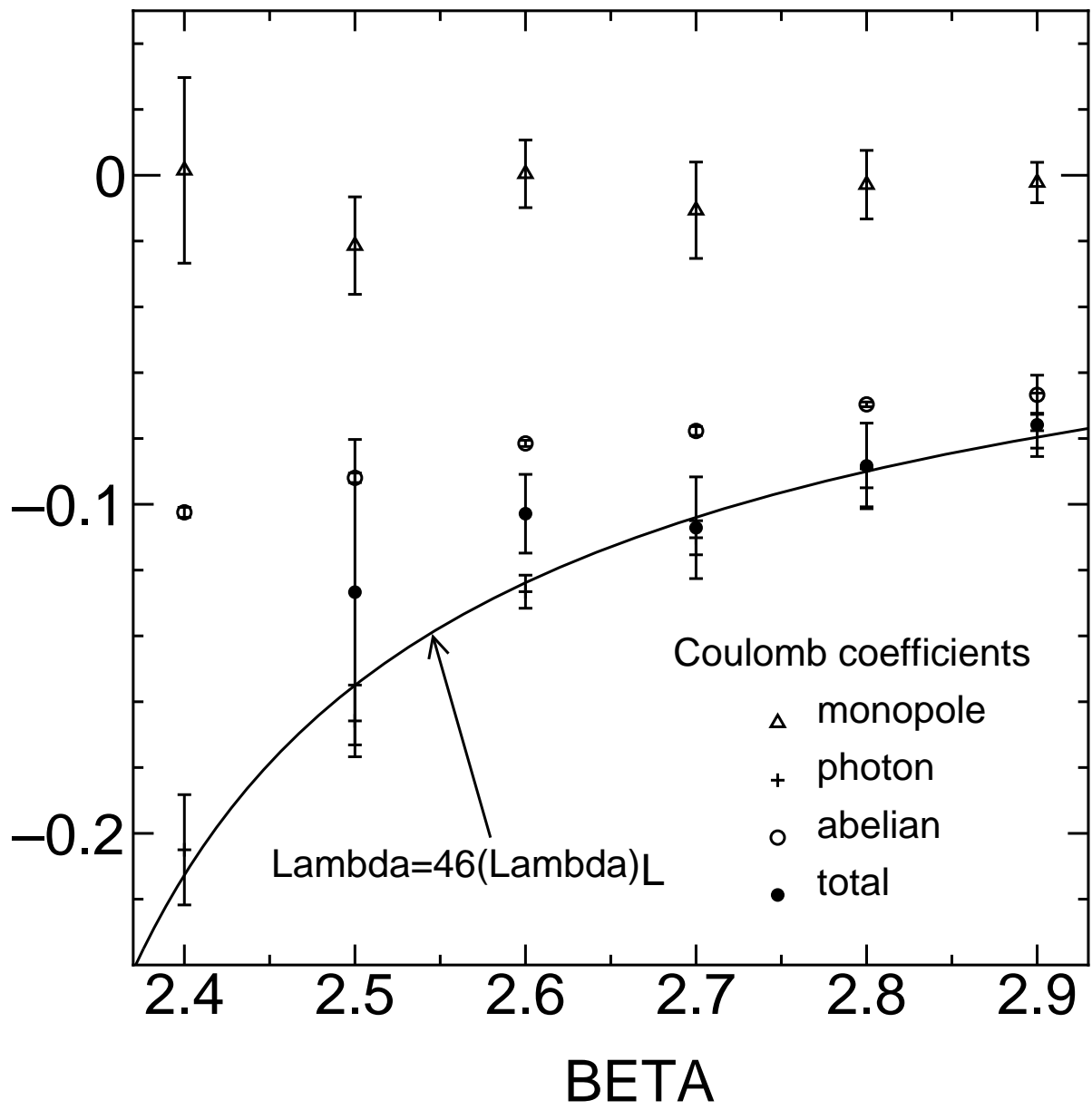
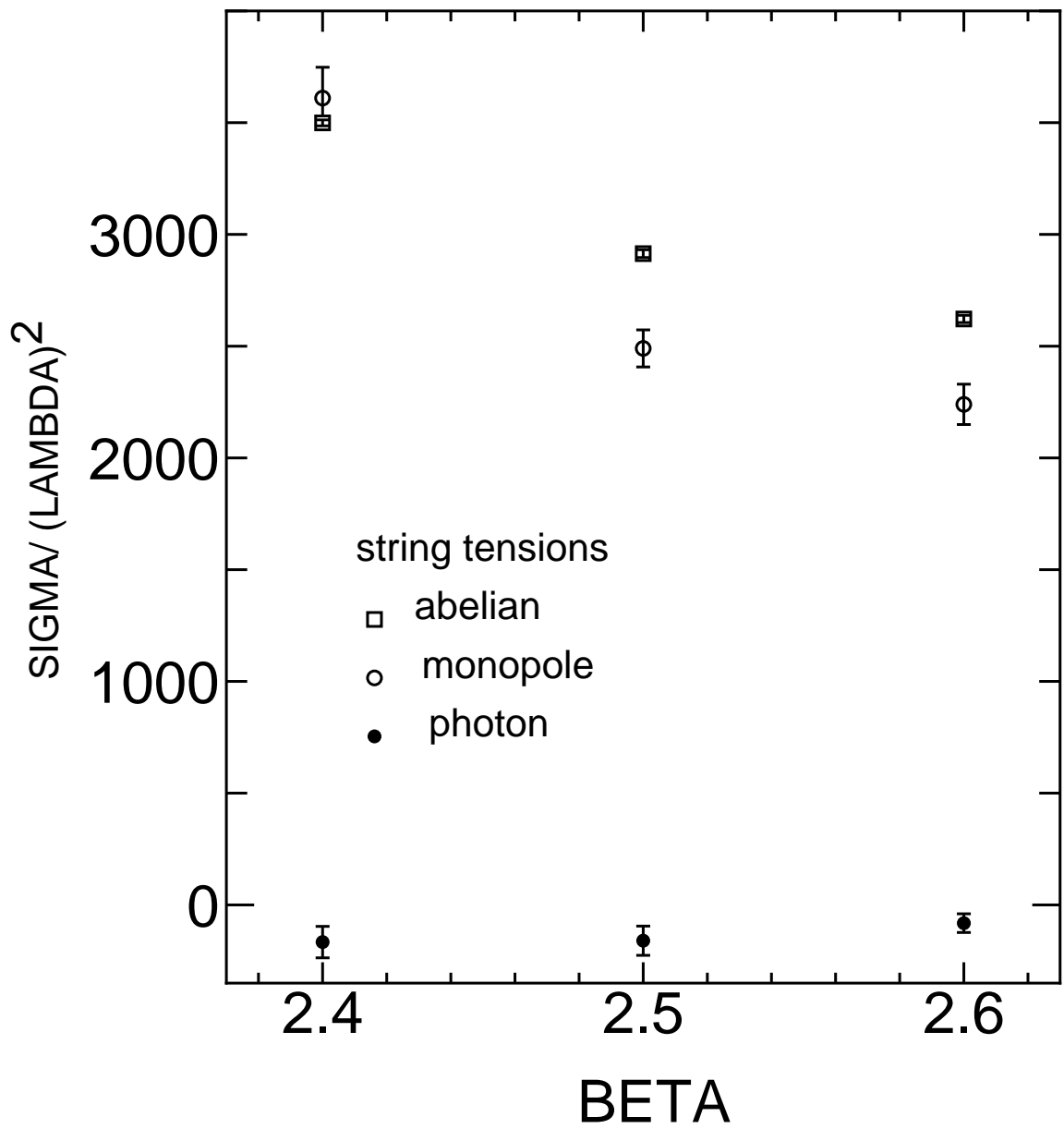
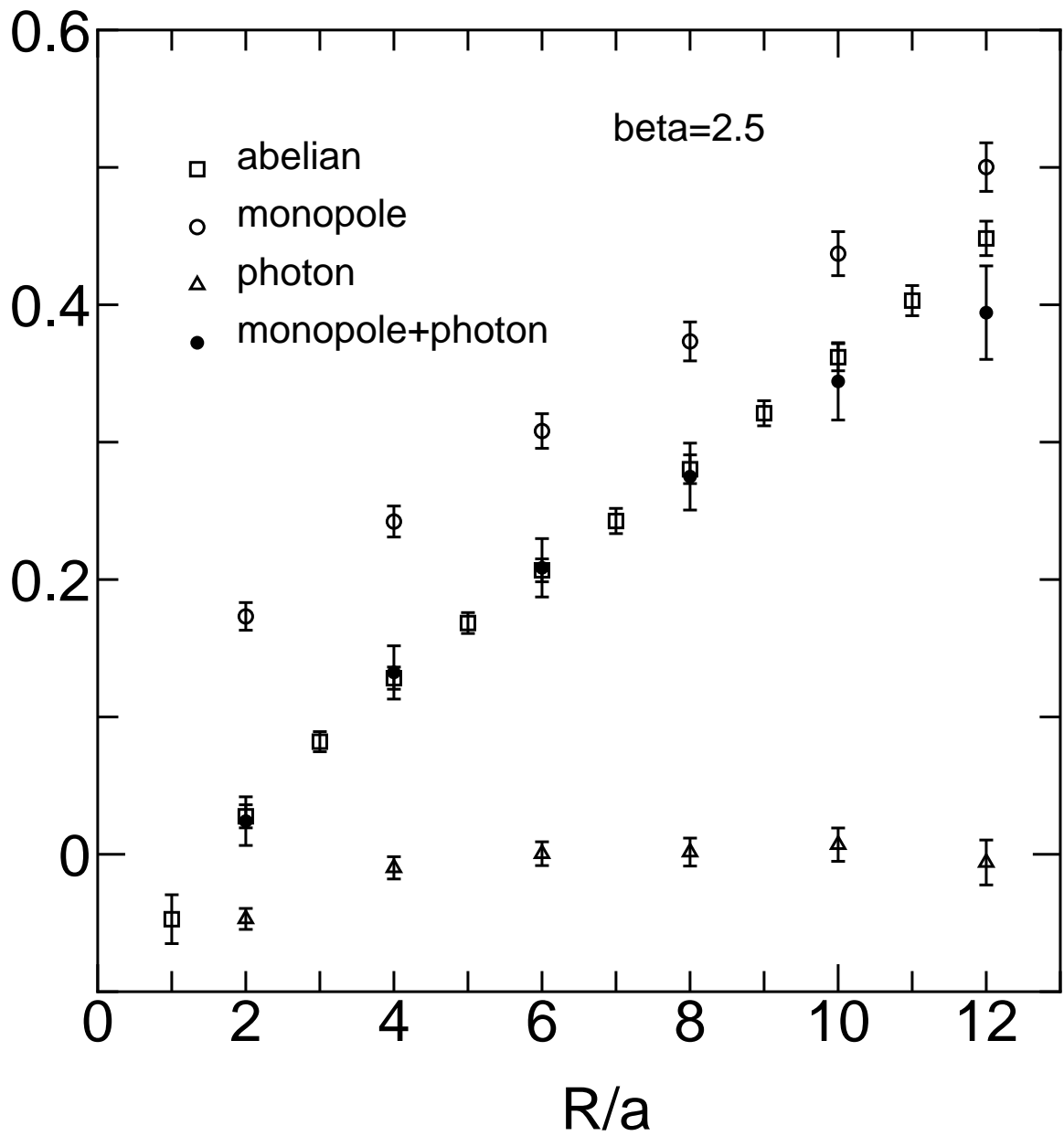
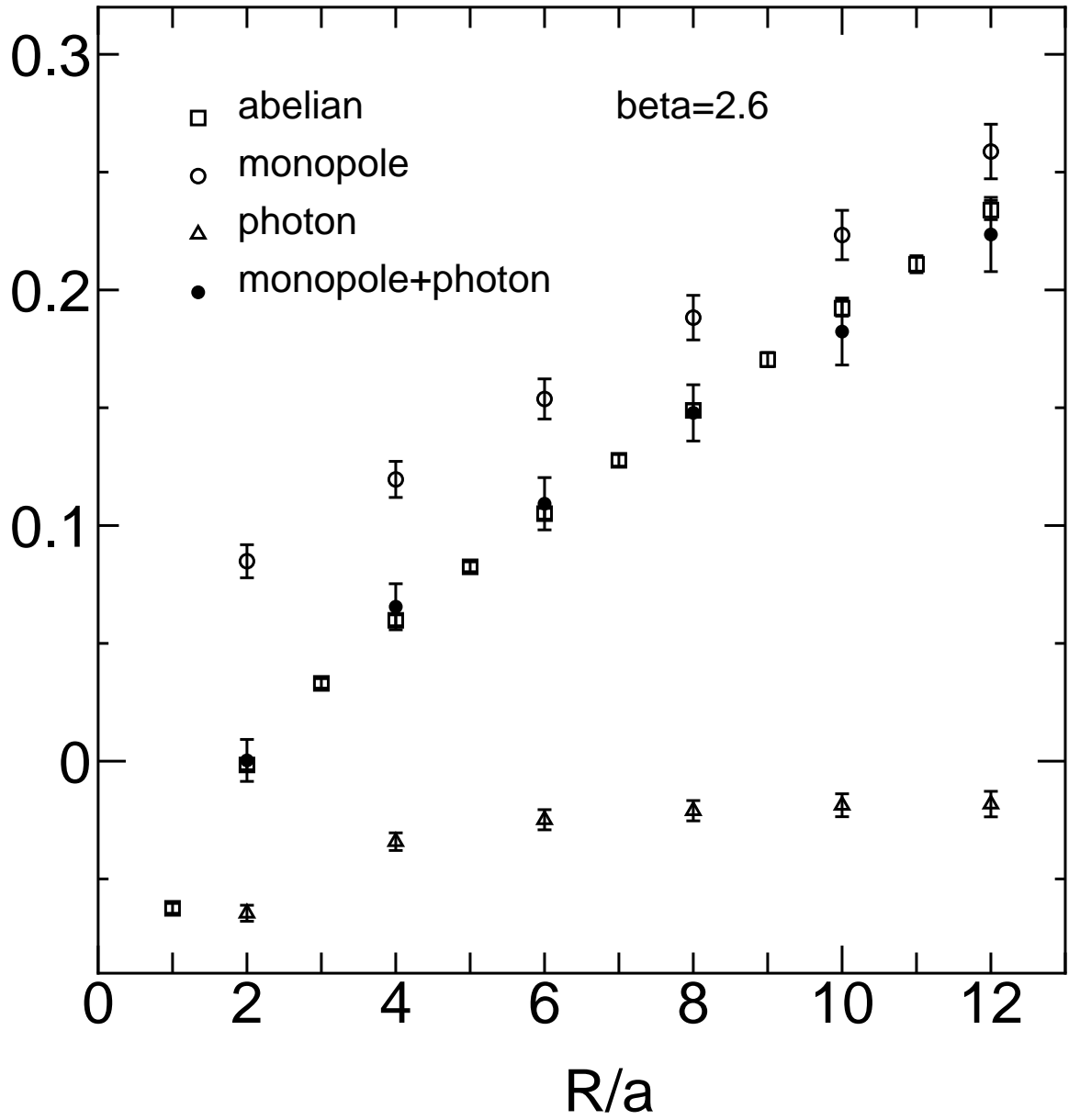
\end{document}